\documentclass{article}

\usepackage[colorlinks,allcolors=blue,pdftex]{hyperref}
\usepackage[english]{babel}
\usepackage[noblocks]{authblk}
\usepackage{booktabs}
\usepackage[square,numbers]{natbib}
\usepackage{float}
\usepackage[export]{adjustbox}

\usepackage[a4paper,margin=2.1cm]{geometry}

\usepackage{amsmath}
\usepackage{amssymb}
\usepackage{graphicx}

\newcommand\keywords[1]{%
    \begingroup
    \let\and\\
    \par
    \noindent\textbf{Keywords:} #1\par
    \endgroup
}

\title{Deep-learning in the bioimaging wild: \\ Handling ambiguous data with \textit{deepflash2}}

\author[1*]{Matthias Griebel}
\author[2]{Dennis Segebarth}
\author[1]{Nikolai Stein}
\author[2]{Nina Schukraft}
\author[2,3]{Philip~Tovote}
\author[4]{Robert Blum}
\author[1]{Christoph M. Flath}

\affil[1]{Department of Business and Economics, University of Würzburg, Germany}
\affil[2]{Institute of Clinical Neurobiology, University Hospital Würzburg, Germany}
\affil[3]{Center for Mental Health, University Hospital Würzburg, Germany}
\affil[4]{Department of Neurology, University Hospital Würzburg, Germany}
\affil[ ]{}
\affil[*]{Corresponding author. Email: \href{mailto:matthias.griebel@uni-wuerzburg.de}{matthias.griebel@uni-wuerzburg.de}}

\date{\vspace{-5ex}}

\begin{document}

\maketitle

\begin{abstract}
We present \textit{deepflash2}, a deep learning solution that facilitates the objective and reliable segmentation of ambiguous bioimages through multi-expert annotations and integrated quality assurance.
Thereby, \textit{deepflash2} addresses typical challenges that arise during training, evaluation, and application of deep learning models in bioimaging.
The tool is embedded in an easy-to-use graphical user interface and offers best-in-class predictive performance for semantic and instance segmentation under economical usage of computational resources.

\end{abstract}


\section{Main}

Partitioning images into meaningful segments (e.g., cells, cellular compartments, or other anatomical structures) is a fundamental task in bioimage analysis as it facilitates subsequent quantification and statistical evaluation of image features. Performing such segmentation tasks manually is tedious and time-consuming.  
Conversely, its automation promises additional insights, more precise analyses, and more rigorous statistics \cite{falk2019u}.
Deep learning (DL) has proven to be a flexible method to analyze large amounts of bioimage data \citep{ronneberger2015u} and numerous solutions for automated segmentation have been proposed \citep{falk2019u, haberl2018cdeep3m, berg2019ilastik, von2021democratising, bannon2021deepcell, isensee2021nnu, stringer2021cellpose, lucas2021open}.
Depending on annotated training data, these tools and analysis pipelines are well suited for settings where the observable phenomena exhibit a high signal-to-noise ratio (SNR), for instance, in monodispersed cell cultures. 
However, the SNR in bioimages is often low, influenced by experimental conditions, sample characteristics, and imaging trade-offs.
Such image material is inherently ambiguous which hampers a reliable analysis.
A case in point is the analysis of fluorescent images of complex brain tissue -- a core technique in modern neuroscience -- which is frequently subject to various sources of ambiguity such as cellular and structural diversity, heterogenous staining conditions and challenging image acquisition processes.

With \textit{deepflash2}, we introduce a DL-based analysis tool for fast and reliable segmentation of ambiguous microscopy images. By integrating annotations from multiple experts and providing quality assurance for the analysis of new images, the tool bridges key challenges during model training, evaluation, and application. 
\textit{Training and evaluation challenges} commence with the manual annotation process. Here, human experts rely on heuristic criteria (e.g., morphology, size, signal intensity) to cope with low SNRs. Relying on a single human expert's annotations for training can result in biased DL models \citep{segebarth2020objectivity}.
At the same time, inter-expert agreement suffers in such settings, which, in turn, leads to ambiguous training annotations \citep{falk2019u,Niedworok2016}. 
Without reliable annotations there is no obvious ground truth, which complicates both model training and evaluation.
The \textit{application challenge} emerges when DL models are deployed for analyzing large numbers of bioimages. This scaling-up step is a crucial leap of faith for users as it effectively means delegating control over the study to a black box system. DL models will generate segmentations for any image. However, the segmentation quality is unknown as the reliability of model generalizations beyond the training data cannot be guaranteed. Selecting a representative subset of images for training and evaluation in a single experiment is already challenging. Maintaining a representative training set across multiple experiments with possibly varying conditions compounds these problems and may eventually prevent reliable automation. For this reason, a viable deployment needs effective quality assurance, or as Ribeiro et al. \citep[p. 1135]{ribeiro2016should} put it ``if the users do not trust [...] a prediction, they will not use it.''

\begin{figure}[!ht]
\includegraphics[center]{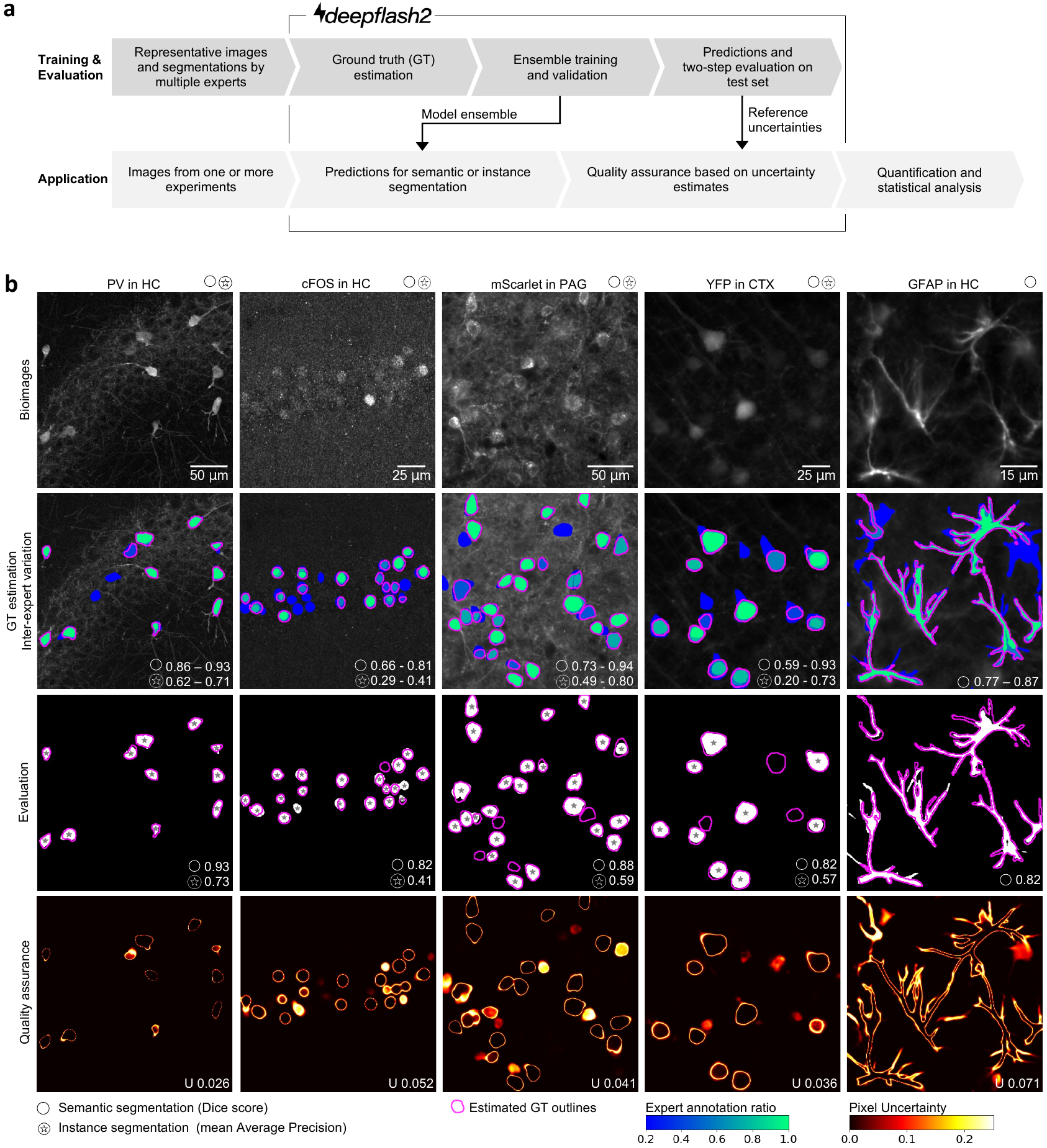}
\caption{\textbf{\textit{deepflash2} pipelines and datasets.} 
\textbf{a}, Proposed integration into the bioimage analysis workflow.
\textbf{b}, Representative image sections from the test sets of five immunofluorescence imaging datasets (first row) with corresponding expert annotations and ground truth (GT) estimation (second row). The inter-expert variation is indicated with ranges (lowest and highest expert similarity to the estimated GT) of the dice score ($DS$) for semantic segmentation and mean Average Precision ($mAP$) for instance segmentation. The predicted segmentations and the similarity to the estimated GT are depicted in the third row, the corresponding uncertainty maps and uncertainty scores $U$ for quality assurance in the fourth row. Areas with a low expert agreement (blue) or differences between the predicted segmentation and the estimated GT typically exhibit high uncertainties. The maximum pixel uncertainty is limited to 0.25.}
\label{fig:fig1}
\end{figure}

\begin{figure}[H]
\includegraphics[center]{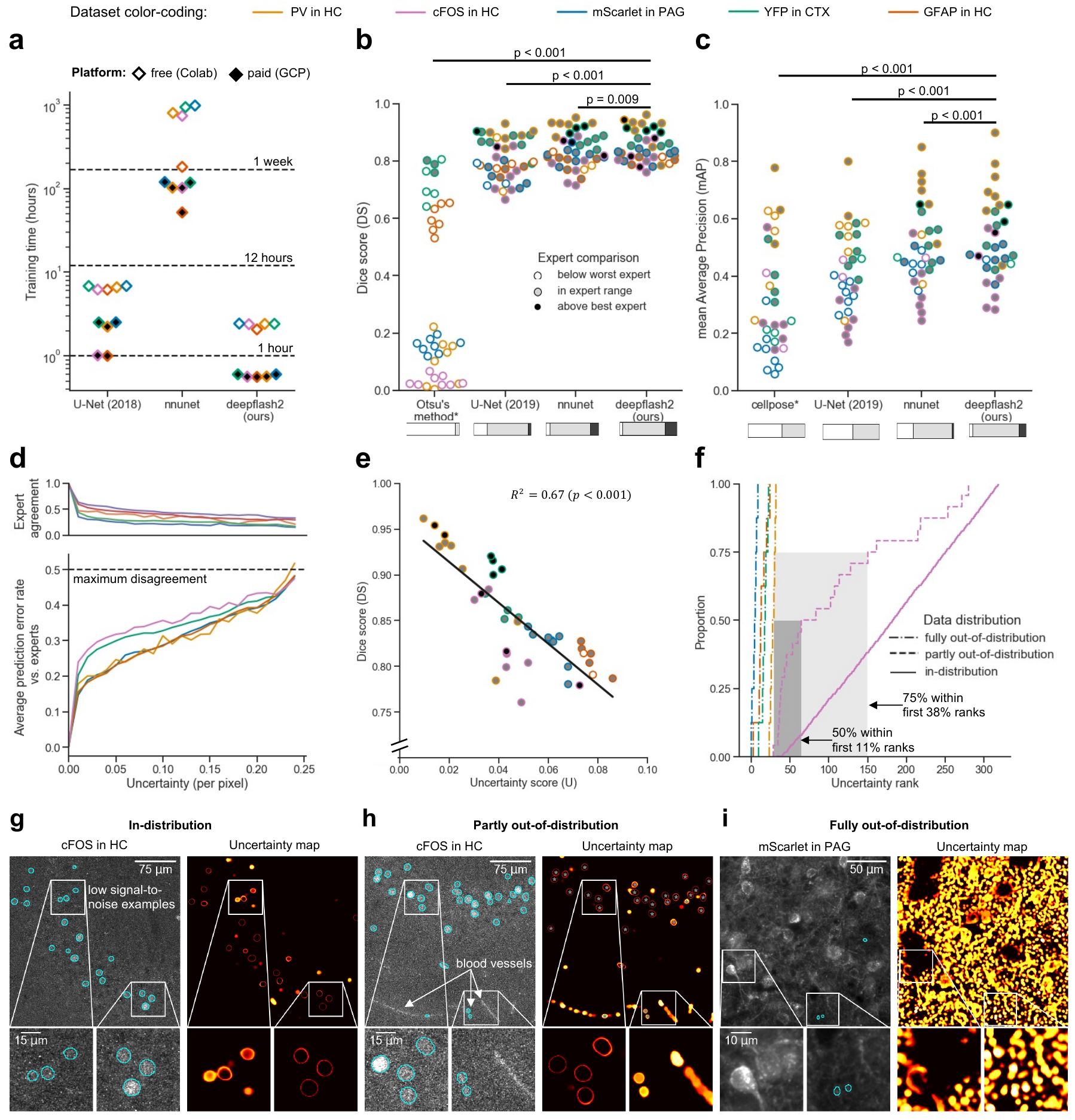}
\caption{
\textbf{Performance comparison and uncertainty evaluation.} 
\textbf{a}, Training durations on different platforms: Google Colaboratory (Colab, free Nvidia Tesla K80 GPU) and Google Cloud Platform (GPC, paid Nvidia A100 GPU).
\textbf{b}, \textbf{c}, Predictive performance on the test sets for (\textbf{b}) semantic segmentation ($N=40$, 8 images for each dataset)  and (\textbf{c}) instance segmentation ($N=32$, 8 images for each depicted dataset except), measured by similarity to the estimated GT. The grayscale filling depicts the comparison against the expert annotation scores. The p-values result from a two-sided Wilcoxon signed-rank test.  *indicates out-of-the-box methods (not fine-tuned to the respective dataset)
\textbf{d}, Relationship between pixel-wise uncertainty and expert agreement (at least one expert with differing annotation; upper plot) and average prediction error rate (relative frequency of deviations between different expert segmentations and the predicted segmentation; lower plot) on the test set.
\textbf{e}, Correlation between dice scores and uncertainties on the test set (marker color codes as in \textbf{b}, \textbf{c}). 
\textbf{f}, Out-of-distribution (\textit{ood}) detection performance using heuristic ranking via uncertainty score. Starting the manual verification of the predictions at the lowest rank all images with deviant fluorescence labels (fully \textit{ood}, $N=32$ images) are detected first. The partly \textit{ood} images (with previously unseen structures, $N=24$) are mostly located in the lower ranks and the in-distribution images (similar to training data of {cFOS in HC}, $N=264$) in the upper ranks.
\textbf{g}, \textbf{h}, \textbf{i}, Representative image crops of the three categories used in \textbf{e}.}
\label{fig:fig2}
\end{figure}

We address these challenges in two consecutive pipelines for \textit{training \& evaluation} and \textit{application} that seamlessly integrate into the bioimage analysis workflow (Fig. \ref{fig:fig1}a).
We illustrate the capabilities of \textit{deepflash2} using five representative fluorescence microscopy datasets of mouse brain tissue with with varying degrees of ambiguity (Fig. \ref{fig:fig1}b, Section \ref{ssec:eval}). We benchmark the tool against other common analysis tools \mbox{(Fig. \ref{fig:fig2}a-c)}, achieving best-in-class predictive performance under economical usage of computational resources. The \textit{deepflash2} pipelines are embedded in a lightweight and easy-to-use graphical user interface (GUI).

Training \& evaluation builds upon a representative sample of the bioimage dataset under analysis, annotated by multiple experts (the annotations can be performed with any tool).
Depending on the biological analysis setting, we distinguish between \textit{semantic} and \textit{instance} segmentation. Semantic segmentation means subdividing the image into meaningful categories \citep{falk2019u}. Instance segmentation further differentiates between multiple instances of the same category by assigning the segmented structures to unique entities (e.g., cell 1, cell 2, ...). 
To derive objective training annotations from multi-annotator data \textit{deepflash2} estimates the ground truth (GT; Section \ref{ssec:gt}) via majority voting or simultaneous truth and performance level estimation (STAPLE) \citep{warfield2004simultaneous}.
\textit{deepflash2} subsequently computes the similarity scores (dice score for semantic segmentation, average precision for instance segmentation; Section \ref{ssec:eval}) between expert segmentations and the estimated GT. These measures of inter-expert variation serve as a proxy for data ambiguity as illustrated in Fig. \ref{fig:fig1}b (first and second row). 
Well-defined fluorescent labels are typically unanimously annotated (green), whereas more ambiguous signals are marked by fewer experts (blue).

DL model training in \textit{deepflash2} capitalizes on model ensembles to ensure high accuracy and reproducibility in the light of data ambiguity \citep{segebarth2020objectivity}. Furthermore, it facilitates reliable uncertainty quantification \citep{lakshminarayanan2017simple}.
To ensure training efficiency \textit{deepflash2} leverages pre-trained models and advanced training strategies (Section \ref{ssec:train}). This approach yields very competitive training durations (Fig. \ref{fig:fig2}a).

The model ensemble then predicts semantic segmentation maps which are evaluated on a hold-out test set (Fig. \ref{fig:fig1}b, third row). If required, \textit{deepflash2} combines these maps with the flow representations of the \textit{cellpose} library \citep{stringer2021cellpose} to derive reliable instance segmentations (i.e., separation of touching objects).
Each segmentation is accompanied by a predictive uncertainty map and the average foreground uncertainty score $U$ (Fig. \ref{fig:fig1}b, fourth row; Section \ref{ssec:pred}). Note that the model ensembles are solely trained on the estimated GT, that is, there is no longer a concept of ambiguous annotations. However, Fig. \ref{fig:fig2}e confirms that the uncertainty maps reliably capture expert disagreement: Low pixel uncertainty is indicative of high expert agreement, whereas high pixel uncertainty arises in settings where experts submitted ambiguous annotations.
To assess the model validity for bioimage analysis, 
\textit{deepflash2} implements the following two-step evaluation process: 

\begin{enumerate}

    \item \textit{Calculate the similarity scores} between the predicted segmentations and the estimated GT on the test set. 
    
    \item \textit{Relate the performance scores to data ambiguity}. The performance scores of individual experts are used to establish the desired performance range. 
    
\end{enumerate}

Across all evaluation datasets, we find that \textit{deepflash2} achieves best-in-class performance vis-a-vis state-of-the-art benchmark tools for both semantic (Fig. \ref{fig:fig2}b) and instance segmentation (Fig. \ref{fig:fig2}c) tasks (detailed analysis in Supplementary Information 2.2). To disentangle the difficulty of the prediction task (driven by data ambiguity) from the predictive performance we scrutinize the ``gross'' performance by comparing it against the underlying expert annotation scores. We find that \textit{deepflash2} reliably achieves human expert performance and in some cases even outperforms the \textit{best} available expert annotation. The $U$ values of the test set serve as a reference for the quality assurance procedure for the application step (Section \ref{ssec:qa}). We find that the uncertainty score $U$ is a strong predictor for the obtained predictive performance as measured by the dice score (Fig. \ref{fig:fig2}d). Consequently, $U$ can be used as a proxy for the expected performance on unlabeled data.

During application, scientists typically aim to analyze a large number of bioimages without ground truth information. 
To establish trust in its predictions, \textit{deepflash2} enables quality assurance in the following manner: First, the predictions are sorted by decreasing uncertainty score. In situations with high uncertainty scores scientists may want to check predictions through manual inspection using the provided uncertainty maps. Also, \textit{deepflash2} facilitates a single click export-import to ImageJ ROIs (regions of interest), with ROIs sorted by uncertainty. 
This quality assurance process helps the user prioritize the review of more ambiguous instances. Moreover, it facilitates the detection of out-of-distribution images, i.e., images that differ from the training data and are thus prone to erroneous predictions. We showcase the out-of-distribution detection on a large bioimage dataset comprising 264 in-distribution images (same properties as training images), 24 partly out-of-distribution images (same properties with previously unseen structures such as blood vessels), and 32 fully out-of-distribution images (different immunofluorescent labels) (Fig. \ref{fig:fig2}g-i).
Using the uncertainty score for sorting, all fully out-of-distribution images are ranked within the first 32 ranks, and most partly out-of-distribution images are ranked within the first 150 ranks (Fig. \ref{fig:fig2}f). A conservative protocol could require scientists to verify all images with an uncertainty score exceeding the reference uncertainty scores (Section \ref{ssec:qa}). Out-of-distribution images may then be excluded from the analysis or annotated for re-training in an active learning manner \citep{gal2017active}.

The GUI of \textit{deepflash2} runs as a web application inside a Jupyter Notebook, the de-facto standard of computational notebooks in the scientific community \citep{perkel2018jupyter}.
\textit{deepflash2} can be installed locally or in cloud environments such as Google Colaboratory (Colab), enabling quick setup in less than a minute and providing free access to graphics processing units (GPUs) for fast training. 
In contrast to other tools that are only optimized for Colab (e.g., \textit{ZeroCostDL4Mic} \citep{von2021democratising}), the \textit{deepflash2} GUI is based on interactive HTML widgets \citep{Kluyver2016jupyter}, providing a clean interface and limiting cognitive overload by leveraging tool-tips and links to the documentation (\href{https://matjesg.github.io/deepflash2/}{https://matjesg.github.io/deepflash2/}).
\textit{deepflash2} allows users to execute all analysis steps directly in the GUI or use the export functionality to Fiji or spreadsheet software for a more detailed analysis.
The GUI is built on top of the \textit{deepflash2 Python API} and leverages established open-source libraries in the \textit{PyTorch} \citep{paszke2019pytorch} ecosystem (Section \ref{sec:methods}).
While our evaluation data focuses on ambiguous fluorescent images, \textit{deepflash2} can be used for 2D images with an arbitrary number of input channels. For example, the \textit{deepflash2 Python API} was successfully used on periodic acid–Schiff stained 3-channel images. The implementation won a Gold Medal and the Innovation Award in a Kaggle data science competition hosted by the HuBMAP consortium.\footnote{\href{https://www.kaggle.com/matjes/hubmap-deepflash2-judge-price}{https://www.kaggle.com/matjes/hubmap-deepflash2-judge-price}}

In summary, \textit{deepflash2} offers an end-to-end integration of DL pipelines for bioimage analysis of ambiguous data. 
An easy-to-use GUI allows researchers without programming experience to rapidly train performant and robust DL model ensembles and monitor their predictions on new data. 
We are confident that \textit{deepflash2} can help establish more objectivity and reproducibility in natural sciences while lowering the overall workload for humans annotators.

\section{Methods}\label{sec:methods}

The \textit{deepflash2} code library is implemented in Python 3, using \textit{numpy}, \textit{scipy}, and \textit{opencv} for the base operations. The ground truth estimation functionalities are based on the \textit{simpleITK} \citep{lowekamp2013design}. The DL related part is built upon the rich ecosystem of \textit{PyTorch} \citep{paszke2019pytorch} libraries, comprising \textit{fastai} \citep{howard2020fastai} for the training procedure, \textit{segmentation models pytorch} \citep{Yakubovskiy_2019} for segmentation architectures, \textit{timm} \citep{rw2019timm} for pre-trained encoders, and \textit{albumentations} \citep{buslaev2020albumentations} for data augmentations.  
Instance segmentation capabilities are complemented using the \textit{cellpose} library \citep{stringer2021cellpose}. The GUI additionally leverages the Jupyter Notebook environment and \textit{ipywidgets} \citep{Kluyver2016jupyter}. 

\subsection{Ground truth estimation}\label{ssec:gt}
To train reproducible and unbiased models \textit{deepflash2} relies on GT estimation from the annotations of multiple experts. \textit{deeepflash2} offers GT estimation via simultaneous truth and performance level estimation (STAPLE) \citep{warfield2004simultaneous} (default in our analyses) or majority voting. This GT estimation is the basis for achieving predictions as close as possible to the unobservable GT. In contrast, recent work on the segmentation of ambiguous data focuses on explicitly modeling the disagreement between the experts \citep{kohl2018probabilistic, ji2021learning}, which is not the main objective of our study. Note that, due to the ambiguities in the data, GT estimation can yield biologically implausible results (e.g., by merging the areas of two cells). We corrected such artifacts in our test sets.

\subsection{Training}\label{ssec:train}
\paragraph{Network Architecture}
Powered by the \textit{segmentation models pytorch} package \citep{Yakubovskiy_2019}, \textit{deepflash2} allows the user to select from various architectures (e.g., Unet  \citep{ronneberger2015u}, Unet++ \citep{zhou2018unet}  or DeepLabV3+ \citep{chen2018encoder}) and a wide range of encoders (e.g., ResNet \citep{he2016deep}, EfficientNet \citep{tan2019efficientnet}, or ResNeSt \citep{zhang2020resnest}). During the development of \textit{deepflash2} we evaluated the performance of different architecture and encoder combinations. We then chose a Unet architecture with a ResNet-34 encoder as a baseline for all experiments of this study. There was no combination that outperformed this baseline on all datasets in a stable training regime. However, we found that switching to more current encoders such as ResNeSt can improve the results on some datasets. 
If available, the encoders can be initialized with pre-trained weights to allow better feature extraction and fast training convergence. The remaining weights are initialized from a truncated normal distribution \citep{he2015delving}. This approach combines the desirable properties of pretraining and random initialization that facilitate diversity in model ensembles. We used imagenet \citep{deng2009imagenet} pre-trained weights in all our experiments.

\paragraph{Training procedure}
Each model is trained using the \textit{fine-tune} policy of the \textit{fastai} library \citep{howard2020fastai}. This entails freezing of the encoder weights, \textit{one-cycle training} \citep{smith2018disciplined} of one epoch, unfreezing the weights, and again \textit{one-cycle-training}. The epochs of the second training cycle depend on the number of training images and are computed such that a fixed number of training iterations is reached.
During each epoch, we sample equally sized patches from each image in the training data.
To address the issue of class imbalances, we use a weighted random sampling approach that ensures that the center points of the patches are sampled equally from each class. This kind of sampling also contributes to the data augmentation pipeline, along with other random augmentations such as rotating, flipping, and gamma correction. Users can adjust these augmentations or add more augmentations (e.g., contrast limited adaptive histogram equalization or grid distortions).
We use the mean of the cross-entropy and Dice loss \citep{drozdzal2016importance} as learning objective. \textit{deepflash2} also provides options for common segmentation loss functions such as Focal \citep{lin2017focal}, Tversky \citep{salehi2017tversky}, or Lovasz \citep{berman2018lovasz}.
We trained each model with 100 iterations in the first (frozen encoder weights) cycle and 2500 iterations in the second cycle using a mini-batch size of four (patch size $512 \times 512$), the Adam optimizer \citep{kingma2014adam} with decoupled weight decay ($0.001$) \citep{loshchilov2017decoupled}, and a base learning rate of $0.001$. The training and validation data for the different models are shuffled by means of a $k$-fold cross-validation (with $k=5$ in our experiments). Users can customize all training settings, for example, by opting for a different optimizer or setting a dataset-specific learning rate using the learning rate finder.

\subsection{Prediction}\label{ssec:pred}
\paragraph{Semantic segmentation}
For the semantic segmentation of a new image with features $\mathbf{X} \in \mathbb{R}^{d \times c}$  \textit{deepflash2} predicts a semantic segmentation map $\mathbf{y} \in\{1, \ldots, K\}^{d}$, with $K$ being the number of classes, $d$ the dimensions of the input and $c$ the input channels. Without loss of generality class $1$ is defined as background.
We use the trained ensemble of $M$ deep neural networks to model the probabilistic predictive distribution $p_{\theta}\left(\mathbf{y} \mid \mathbf{X} \right)$, where $\theta = (\theta_1, \dots, \theta_M)$ are the parameters of the ensemble.
Here, we leverage a sliding window approach with overlapping borders and Gaussian importance weighting \citep{isensee2021nnu}.
We improve the prediction accuracy and robustness using $T$ deterministic test-time augmentations (rotating and flipping the input image).
Each augmentation $t \in \{1,...,T\}$ applied to an input image creates an augmented feature matrix $\mathbf{X_t}$.
To combine all predictions we follow Lakshminarayanan et al. \cite{lakshminarayanan2017simple} and treat the ensemble as a uniformly-weighted mixture model to derive

\begin{equation}
p (\mathbf{y} \mid \mathbf{X}) = \frac{1}{T} \sum_{t=1}^{T} \frac{1}{M} \sum_{m=1}^{M} p_{\theta_{m}}\left(\mathbf{y} \mid \mathbf{X_t}, \theta_{m}\right)
\end{equation}
with $p_{\theta_{m}}\left(\mathbf{y} \mid \mathbf{X_t}, \theta_{m} \right)=\operatorname{Softmax}\left(\mathbf{f}_{\theta_{m}}(\mathbf{X_t})\right)$ and $\mathbf{f}_{\theta_{m}}$ representing the neural network parametrized with $\theta_m$. We use $M=5$ models and $T=4$ augmentations in all our experiments. Finally, we obtain the predicted segmentation map  
\begin{equation}
\hat{\mathbf{y}}=\underset{\mathbf{k} \in \{1,\dots,K\}^d}{\arg \max}\; p (\mathbf{y}=\mathbf{k} \mid \mathbf{X}).
\end{equation}

\begin{sloppypar}

\paragraph{Uncertainty quantification}
The uncertainty is typically categorized into \textit{aleatoric} (statistical or per-measurement) uncertainty and \textit{epistemic} (systematic or model) uncertainty \citep{der2009aleatory}. 
To approximate the uncertainty maps of the the predicted segmentations we follow the approach of Kwon et al. \cite{kwon2020uncertainty}. Here, we replace the Monte-Carlo dropout approach of \cite{gal2016dropout} with deep ensembles, which have proven to produce well-calibrated uncertainty estimates and a more robust out-of-distribution detection \citep{lakshminarayanan2017simple}. In combination with  test-time augmentations (inspired by Wang et al. \cite{wang2019aleatoric}) we approximate the predictive (hybrid) uncertainty the for each class $k \in \{1,\dots,K\}$ as

\begin{equation}
\begin{aligned}
\operatorname{Var}_{p\left(\mathbf{y}=k \mid \mathbf{X} \right)} \approx &  \underbrace{\frac{1}{T} \sum_{t=1}^{T}  \frac{1}{M} \sum_{m=1}^{M}  \left[p_{\theta_{m}}\left(\mathbf{y}=k \mid \mathbf{X}_t, \theta_{m}\right)-p_{\theta_{m}}\left(\mathbf{y}=k \mid \mathbf{X}_t, \theta_{m}\right)^{2}\right]}_{\text {epistemic uncertainty}} \\ & +
 \underbrace{\frac{1}{T} \sum_{t=1}^{T}  \frac{1}{M} \sum_{m=1}^{M} \left[p_{\theta_{m}}\left(\mathbf{y}=k \mid \mathbf{X}_t, \theta_{m}\right)-p \left(\mathbf{y}=k \mid \mathbf{X}\right)\right]^{2}}_{\text {aleatoric uncertainty}}
 \end{aligned}
\end{equation}
where $p\left(\mathbf{y}=k \mid \mathbf{X} \right)$ denotes probabilities of a single class $k$.
\end{sloppypar}

To allow an intuitive visualization and efficient calculation in multi-class settings, we aggregate the results of the single classes to retrieve the final predictive uncertainty map:
\begin{equation}
\operatorname{Var}_{p\left(\mathbf{y} \mid \mathbf{X},\theta \right)} = \frac{1}{K} \sum_{k=1}^{K}   \operatorname{Var}_{p\left(\mathbf{y}=k \mid \mathbf{X},\theta \right)}
\end{equation}

Note that, due to symmetries, the results may only differ from the general formulation in \cite{kwon2020uncertainty} for $K>2$.

For the heuristic sorting and out-of-distribution detection we define an aggregated uncertainty metric on image level.
Let $\hat{y}_{i}$ be the predicted segmentation of pixel $i$, $\mathbf{x}_i$ the feature vector of pixel $i$ and $N$ the total number of pixels defined by $d$. We define the scalar valued foreground uncertainty score for all predicted $N_f = \Big\{i \in \left\{1,...,N\right\} \mid \hat{y}_{i} > 1\Big\}$ as

\begin{equation}
    U_{p\left(\mathbf{y} \mid \mathbf{X},\theta \right)} := \frac{1}{|N_f|}\sum_{i \in N_f} \operatorname{Var}_{p\left(y_{i} \mid \mathbf{x}_i,\theta \right)}.
\end{equation}

\paragraph{Instance segmentation}
If the segmented image contains touching objects (e.g., cells that are in close proximity), \textit{deepflash2} offers an option for reliable instance segmentation. For this, we use the combined predictions of each class $p\left(\mathbf{y}=k \mid \mathbf{X}\right)$ to predict the flow representations using the \textit{cellpose} library \citep{stringer2021cellpose}. We then leverage the post-processing pipeline of \textit{cellpose} to derive instance segmentations by combining the flow representations with the predicted segmentation maps $\hat{y}$. This procedure scales to an arbitrary number of classes and is, in contrast to the original \textit{cellpose} implementation, not limited to one (or two) of input channels.

\subsection{Evaluation}\label{ssec:eval}

\paragraph{Evaluation metrics} 
For semantic segmentation, we calculate the similarity of two segmentation masks $\mathbf{y}_{a}$ and $\mathbf{y}_{b}$ using the dice score. For binary masks, this metric is defined as

\begin{equation}
DS:=\frac{2TP}{2 TP+FP+FN},
\end{equation}
\\
where the \textit{true positives} ($TP$) are the sum of all matching positive (pixels) elements of $\mathbf{y}_{a}$ and $\mathbf{y}_{b}$, and the \textit{false positives} ($FP$) and \textit{false negatives} ($FN$) the sum of positive elements that only appear in $\mathbf{y}_{a}$ or $\mathbf{y}_{b}$, respectively.
In multi-class settings we use \textit{macro} averaging, i.e., we calculate the metrics for each class and then find their unweighted mean.
The dice score is commonly used for semantic segmentation tasks but is unaware of different instances (sets of pixels belonging to a class and instance). 

For instance segmentation, let $\mathbf{y}_{a}^{I}$ and $\mathbf{y}_{b}^{I}$ be two instance segmentation masks that contain a finite number of instances $I_a$ and $I_b$, respectively. An instance $I_a$ is considered a match (\textit{true positive} -- $TP_{\eta}$) if an instance $I_b$ exists with an \textit{Intersection of Union} (also known as \textit{Jaccard index})
$IoU(I_a, I_b)=\frac{I_a \cap I_b}{I_a \cup I_b}$ exceeding a threshold $\eta \in [0,1]$. Unmatched instances $I_a$ are considered as \textit{false positives} ($FP_{\eta}$), unmatched instances $I_b$ as \textit{false negatives} ($FN_{\eta}$). We define the Average Precision at a fixed threshold $\eta$ as $AP_{\eta}:=\frac{T P_{\eta}}{T P_{\eta}+F N_{\eta}+F P_{\eta}}$. To become independent of fixed values for $\eta$ it is common to average the results over different $\eta$. The resulting metric is known as \textit{mean Average Precision} and defined as

\begin{equation}
mAP :=\frac{1}{|H|} \sum_{\eta \in H} AP_{\eta}.
\end{equation}

We use a set of $10$ thresholds $H=\{\eta \in [0.50, \dots, 0.95] \mid \eta \equiv 0 \mod{0.05}\}$ for all evaluations. This corresponds to the metric used in the COCO object detection challenge \citep{lin2014microsoft}. Additionally, we exclude all instances $I$ that are below a biologically viable size from the analysis. The minimum size is derived from the smallest area annotated by a human expert: 61 pixel (\textit{PV in HC}), 30 pixel (\textit{cFOS in HC}),  385 pixel (\textit{mScarlet in PAG}), and 193 pixel (\textit{YFP in CTX}).
    
\paragraph{Evaluation datasets}

We evaluate our pipeline on five datasets that represent common bioimage analysis settings. The datasets exemplify a range of fluorescently labeled (sub-)cellular targets in mouse brain tissue with varying degrees of data ambiguity. Detailed methodological  information on each dataset, including sample preparation and image acquisition for datasets that were generated for this study, can be found in the Supplementary Information S1. 

\begin{sloppypar}
The \textit{PV in HC} dataset published by Segebarth et al. \cite{segebarth2020objectivity} describes indirect immunofluorescence labeling of Parvalbumin-positive (PV-positive) interneurons in the hippocampus.
Morphological features are widely ramified axons projecting to neighbored neurons for soma-near inhibition of excitatory neuronal activity \cite{hu2014fast}. The axonal projections densely wrap around the somata of target cells. This occasionally causes data ambiguities when somata of the PV-positive neurons need to be separated from the PV-positive immunofluorescent signal around the soma of neighboured cells. Thresholding approaches such as Otsu’s method (see Fig. \ref{fig:fig2}b) typically fail at this task as it requires to differentiate between rather brightly labeled somata that express PV in the cytosol vs. brightly labeled PV-positive axon bundles that can appear in the neighborhood.
\end{sloppypar}

The publicly available \textit{cFOS in HC} dataset \citep{segebarth2020dataset} 
describes indirect immunofluorescent labeling of the transcription factor cFOS in different subregions of the hippocampus after behavioral testing of the mice \cite{segebarth2020objectivity}. The counting or segmentation of cFOS-positive nuclei is an often used experimental paradigm in the neurosciences. The staining is used to investigate information processing in neural circuits \cite{ruediger2011learning}. The low SNR of cFOS labels for most but not all image features, renders its heuristic 
segmentation a very challenging task. This results in a very high inter-expert variability after manual segmentation (see \citep{segebarth2020objectivity} and Supplementary Information Fig. S2.1). We use 288 additional images of this dataset to demonstrate the out-of-distribution detection capabilities of \textit{deepflash2}. There are no expert annotations available for the additional images, however, 24 images comprise characteristics that do not occur in the training data. Such partly out-of-distribution images exhibit elements that may distort the analysis, e.g., blood vessels, folded tissue, and fluorescent particles (see Supplementary Information Fig. S1.1).

The optogenetic \textit{mScarlet in PAG} dataset shows an indirect immunofluorescent post-labeling of the red-fluorescent protein mScarlet, after viral expression in the peri-aqueductal gray. Here, microscopy images visualize mScarlet, tagged to the light-sensitive inhibitory opsin OPN3. The recombinant protein was delivered via stereotactic injection of an adeno-associated viral vector to the PAG. Recombinant opsins are often used for optogenetics, a key technology in the neurosciences \citep{rost2017optogenetic}. This allows control of neuronal activity in selected neuron populations. The method needs to be controlled by immunofluorescence microscopy to verify the quantitative expression and precise localization of the optogenetic proteins. Due to a plethora of factors (e.g. virus injection, virus titer at the locus of injection, experiment-specific success), the resulting (sub-)cellular distribution and intensity of the fluorescent signals results in highly ambiguous images, such as those depicted in Fig. \ref{fig:fig1}b. This renders the segmentation of infected neurons in optogenetics challenging and tedious \cite{falk2019u}.

The \textit{YFP in CTX} dataset shows direct fluorescence of yellow fluorescent protein (YFP) in the cortex of so-called thy1-YFP mice. In thy1-YFP mice, a fluorescent protein is expressed in the cytosol of neuronal subtypes with the help of promoter elements from the thy1 gene \cite{feng2000imaging}. This provides a fluorescent Golgi-like vital stain that can be used to investigate disease-related changes in neuron numbers or neuron morphology, for instance for hypothesis-generating research in neurodegenerative diseases (e.g. Alzheimers disease). Here, computational bioimage analysis is aggravated by the pure intensity of the label that causes strong background signals by light scattering or out-of-focus light. Both can blur the signal borders in the image plane.

Finally, the \textit{GFAP in HC} dataset shows indirect immunofluorescence signals of glial acidic fibrillary protein (GFAP) in the hippocampus. Anti-GFAP labeling is one of the most commonly used stainings in the neurosciences and is also used for histological examination of brain tumor tissue. Glial cells labeled by GFAP in the hippocampus show different morphologies (e.g. radial-like or star-like). GFAP-positive cells occupy separate anatomical parts \cite{bushong2002protoplasmic} (like balls in a ball bath). Thus, it is highly laborious to manually segment the spatial area of GFAP-positive single astrocytes in a brain slice. 
Here, the extensions of the GFAP-labeled astrocytic skeleton cannot be separated from parts of neighboring astrocytes, rendering a reliable instance separation and thus instance segmentation impossible. Albeit the signal is typically bright and very clear around the center of the cell, the signal borders of the radial fibers become ambiguous due to the 3D-ball-like structure, low SNR at the end of the fibers, and out-of-focus light interference. 

A high-level comparison of the key dataset characteristics is provided in Table \ref{tab:datasets}.

\begin{table}[H]
	\caption{Comparison of datasets}
	\centering
	\scalebox{.85}{
	\begin{tabular}{lrrrrr}
		\toprule
		   & PV in HC \citep{segebarth2020objectivity}& cFOS in HC \citep{segebarth2020objectivity} & mScarlet in PAG & YFP in CTX & GFAP in HC \\
		\midrule
		Annotation target & somata & nuclei & somata & somata  & morphology \\ 
		Semantic segmentation & yes  & yes & yes & yes  & yes   \\
		Instance segmentation & yes  & yes & yes & yes  & no   \\
		Train images & 36  & 36 & 20 & 20  & 20   \\
		Test images  & 8 & 8 & 8 & 8 & 8 \\
		Experts     & 5 & 5 & 4-5 & 4-5  & 3 \\
		Additional images  & -- & 288 & -- & -- &  -- \\
		Fluorescence Microsc. & confocal & confocal & light& light  & light \\
		Size (pixel) & 1024$\times$1024 & 1024$\times$1024 & 2752$\times$2208 &  2752$\times$2208  &  580$\times$580 \\
		Resolution (px/$\mu$m) & 1.61  & 1.61 & 3.7 & 3.7 & 3.7   \\
		\bottomrule
	\end{tabular}
	}
	\label{tab:datasets}
\end{table}


\paragraph{Performance benchmarks}
We benchmark the predictive performance of \textit{deepflash2} against a select group of well-established algorithms and tools. These comprise the U-Net of \citep{falk2019u} and \textit{nnunet} \citep{isensee2021nnu} for both semantic and instance segmentation as well as two out-of-the-box baselines. We utilize Otsu's method \citep{otsu1979threshold} as a simple baseline for semantic segmentation and \textit{cellpose} \citep{stringer2021cellpose} as a generic baseline for (cell) instance segmentation. Additionally, we benchmark \textit{deepflash2} against fine-tuned \textit{cellpose} models and ensembles, showing superior performance of our method (see Supplementary Information Table S2.1). \textit{cellpose} has previously proven to outperform other well-known methods for instance segmentation (e.g., Mask-RCNN \citep{he2017mask} or StarDist \citep{schmidt2018cell}).

For each dataset, we apply the tools as described by their developers to render the comparison as fair as possible. We train the U-Net of \citep{falk2019u} on a 90/10 train-validation-split for 10,000 iterations (learning rate of 0.00001 and the Adam optimizer \citep{kingma2014adam}) using the authors' \textit{TensorFlow 1.x} implementation. This includes all relevant features such as overlapping tile strategy and border-aware loss function. We derive the parameter values for the loss function (\textit{border weight factor} ($\lambda$), \textit{border weight sigma} ($\sigma_{sep}$), and \textit{foreground-background-ratio} ($\nu_{bal}$) by means of Bayesian hyperparameter tuning: Parv in HC: $\lambda=25, \sigma_{sep}=10, \nu_{bal}=0.66$; 
cFOS in HC: $\lambda=44, \sigma_{sep}=2, \nu_{bal}=0.23$;  
mScarlet in PAG: $\lambda=15, \sigma_{sep}=10, \nu_{bal}=0.66$; 
YFP in CTX: $\lambda=15, \sigma_{sep}=5, \nu_{bal}=0.85$; 
GFAP in HC: $\lambda=1, \sigma_{sep}=1, \nu_{bal}=0.85$.

We train the self-configuring \textit{nnunet} (version 1.6.6) model ensemble \citep{isensee2021nnu} following the authors' instructions provided on GitHub. 

\textit{cellpose} provides three pretrained model ensembles (\textit{nuclei}, \textit{cyto}, and \textit{cyto2}) for the out-of-the-box usage \citep{stringer2021cellpose}. We select the ensemble with the highest score on the training data: \textit{cyto} for Parv in HC and YFP in CTX; \textit{cyto2} for cFOS in HC and mScarlet in PAG. During inference we fix the cell diameter (in pixel) for each dataset:   
Parv in HC: $24$; 
cFOS in HC: $15$;  
mScarlet in PAG: $55$; 
YFP in CTX: $50$. 
We additionally provide a performance comparison for fine-tune \textit{cellpose} models and ensembles in the Supplementary Information S2.2. We use the \textit{cellpose} GitHub version with commit hash 316927e (August 26, 2021) for our experiments. 

We repeat our experiments with different seeds to ensure that our results are robust and reproducible (see Supplementary Information S2.2).
The experiments for training duration comparison are executed on the \textit{free} platform Google Colaboratory (Nvidia Tesla K80 GPU, 2 vCPUs; times were extrapolated when the 12-hour limit was reached) and the \textit{paid} Google Cloud Platform (Nvidia A100 GPU, 12 vCPUs). The remaining experiments are executed locally (Nvidia GeForce RTX 3090) or in the cloud (Google Cloud Platform on Nvidia Tesla K40 GPUs).

\subsection{Quality Assurance}\label{ssec:qa}

Once the \textit{deepflash2} model ensemble is deployed for predictions on new data, the quality assurance process helps the user prioritize the review of more ambiguous or out-of-distribution images. The predictions on such images are typically error-prone and exhibit a higher uncertainty score $U$. Thus, \textit{deepflash2} automatically sorts the predictions by decreasing uncertainty score. 
Depending on the ambiguities in the data and the expected prediction quality (inferred from the hold-out test set), 
a conservative protocol could require scientists to verify all images with an uncertainty score exceeding a threshold $ U_{min}$. Given the the hold-out test set $Q = \left\{(\mathbf{X}_1, \mathbf{y}_1),\dots,(\mathbf{X}_L, \mathbf{y}_L) \right\}$ where L is the number of samples we define 

\begin{equation}
     U_{min}:=\min \left\{U_{p\left(\mathbf{y} \mid \mathbf{X},\theta \right)} \mid (\mathbf{y}, \mathbf{X}) \in Q,  S(y,\hat{y})<\tau \right\}.
\end{equation}

with $S(y,\hat{y})$ being an arbitrary evaluation metric (e.g., $DS$ or $mAP$) and $\tau \in [0,1]$ a threshold that satisfies the prediction quality requirements. 
From a practical perspective, this means selecting all predictions from the test set with a score below the pre-defined threshold (e.g., $DS=0.8$) and taking their minimum uncertainty score value $U$ as $ U_{min}$. The verification process of a single image is simplified by the uncertainty maps that allow the user to quickly find difficult or ambiguous areas within the image. 

\section{Code and data availability}

The source code for \textit{deepflash2} is publicly available on GitHub (\href{https://github.com/matjesg/deepflash2}{https://github.com/matjesg/deepflash2}).
The \textit{cFOS in HC} dataset from Segebarth et al. \cite{segebarth2020objectivity} is publicly available and can be downloaded from Dryad \citep{segebarth2020dataset}.
The remaining data will be made available upon final acceptance of the paper.

\begin{sloppypar}
\bibliography{main.bib}
\end{sloppypar}

\end{document}


\section*{Supplementary Information}

This document contains supplementary information for the manuscript ``Deep-learning in the bioimaging wild: Handling ambiguous data with \textit{deepflash2}''.

\section{Extended dataset descriptions}\label{sec:datasets}

We use five representative immunofluorescence microscopy datasets (Section 2.4, Table 1 in the main paper) to illustrate the capabilities of \textit{deepflash2} and benchmark its performance against other common analysis tools.

\subsection{Public datasets and data from previous studies}
\begin{figure}[h!]
\begin{center}
\includegraphics[width=1\linewidth]{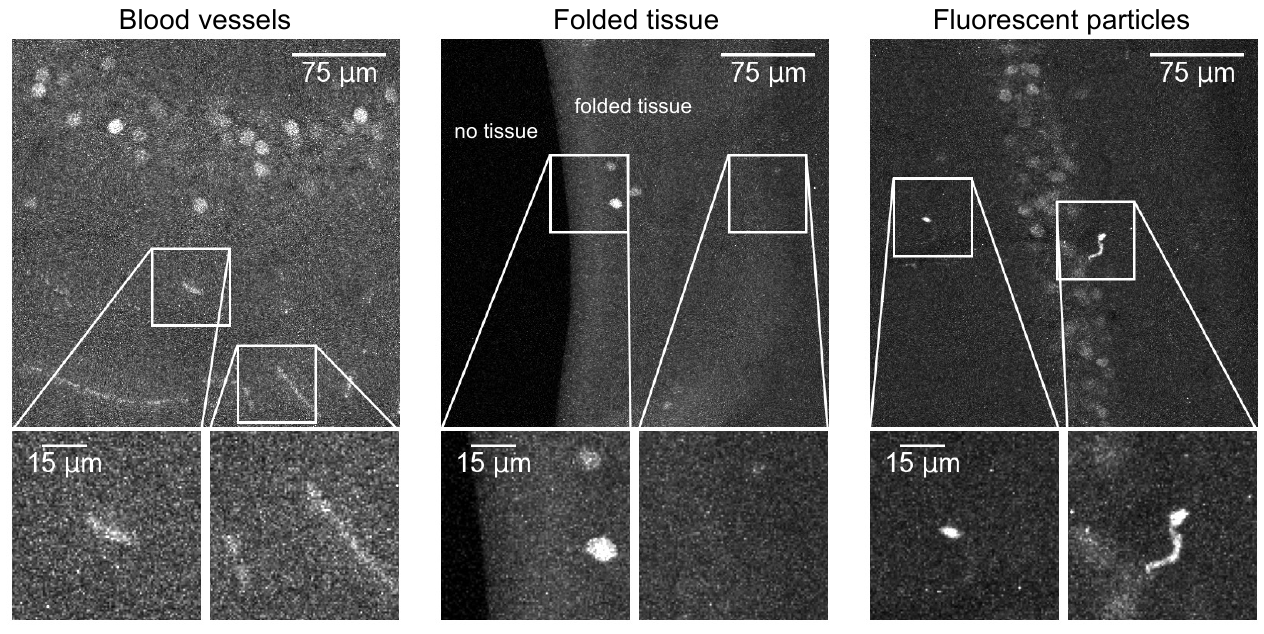}
\caption{\textbf{Partly out-of-distribution images.} Image crops and zoom-ins of three error categories in the extended \textit{cFOS in HC} dataset. 
}
\label{fig:ood_plain}
\end{center}
\end{figure}

The datasets \textit{Parv in HC} and \textit{cFOS in HC} are from the study of Segebarth et al. \citep{segebarth2020objectivity}. The train and test data for \textit{cFOS in HC} are publicly available and can be downloaded from Dryad \cite{segebarth2020dataset}. 
We removed one image (id 1608) from the test sets to ensure a balanced evaluation (eight test images across all five datasets in this study). 
\paragraph{Out-of-distribution detection}
To demonstrate the quality assurance and out-of-distribution detection capabilities of \textit{deepflash2} we used 288 \textit{cFOS in HC} images that were acquired for the study of Segebarth et al. \citep{segebarth2020objectivity}. There are no expert annotations available for these images, however, 24 of these images were excluded from the bioimage analysis of \citep{segebarth2020objectivity} because they did not meet the quality criteria. We classified these \textit{partly out-of-distribution} images into three different error categories for our study: \textit{blood vessels}, if the images contained blood vessels (13 images); \textit{folded tissue} (4 images); \textit{fluorescent particles}, if there was at least one strongly fluorescent particle, unrelated to the actual fluorescent label (7 images). Representative samples of these categories are depicted in Fig. \ref{fig:ood_plain}.

\subsection{Datasets for this study}
The datasets \textit{mScarlet in PAG}, \textit{YFP in CTX}, \textit{GFAP in HC}  were acquired for this study.

\subsubsection{Experimental animals}
All mice used in this study were bred in the animal facility of the Institute of Clinical Neurobiology, at the University Hospital of Würzburg, Germany, and housed under standard conditions (55 $\pm$ 5 \% humidity, 21 $\pm$ 1~\textcelsius, 12:12 hours light:dark cycle) with access to food and water \textit{ad libidum}. VGlut2-IRES-Cre knock-in mice \citep{vong2011leptin} (stock no. 208863), as well as Thy1-YFP mice \citep{feng2000imaging} (stock no. 003782) were obtained from Jackson Laboratory. 
All experiments and experimental procedures were in accordance with the guidelines set by the European Union and our local veterinary authority (Veterinäramt der Stadt Würzburg). In addition, all experiments and experimental procedures were approved by our institutional Animal Care, the Utilization Committee, and the Regierung von Unterfranken, Würzburg, Germany (License numbers: 55.2–2531.01-95/13 and 55.2.2-352-2-509).

\subsubsection{Surgeries}
Male VGlut2-IRES-Cre knock-in mice were injected at an age of four months and adeno-associated virus (AAV) were used as vectors to deliver genetic material into the brain. AAV vectors encoding Cre-dependently for the inhibitory opsin mScarlet \citep{mahn2021efficient} were injected into the periaqueductal gray (PAG) bilaterally. The construct \textit{EF1$\alpha$-DIO-emScarlet-ts-mScarlet-ER} was kindly provided by Simon Wiegert, Center for Molecular Neurobiology Hamburg, Germany. Respective AAV vectors were produced in house (AAV2/5 capsid). 
For stereotactic surgeries, animals were prepared with an administration od Buprenorphin (Buprenorvet, Bayer). Mice were deeply anesthesized with 4-5 \% isoflurane/O2 (Anesthetic Vaporizer, Harvard Apparatus). Animals were fixed into the stereotactic frame (Kopf, Model 1900) and anesthesia was maintained with 1.5-2 \% isoflurane/O2. A subcutaneous injection of Ropivacain (Naropin Aspen) was used for local analgesia before opening the scalp. Craniotomies were performed at bregma coordinates AP -4.5 mm, ML +/- 0.6 mm. A glass pipette (Drummond Scentific) filled with the viral vector and lowered to the target depth of -2.9 mm from bregma. A volume of ~ 100~nl was injected with a pressure injector (NPI electronic). After injection, pipette was hold in place for 8 minutes before retracting. The wound was closed and animal was treated with a subcutaneous injection of Metacam (Metacam, Boehringer Ingelheim) for post-surgery analgesia. After six weeks of expression time, animals were perfused and brain tissue was dissected for further analysis. 

\subsubsection{Sample preparation}
Following intraperitoneal injection (for \textit{YFP in CTX} and \textit{GFAP in HC}): 12 $\mu$l / g bodyweight of a mixture of ketamine (100 mg/kg; Ursotamin, Serumwerk) and xylazine (16 mg/kg; cp-Pharma, Xylavet, Burgdorf, Germany); for \textit{mScarlet in PAG}: urethane (2g/kg; Sigma-Aldrich) at a volume of 200 µl diluted in 0.9\% sterile sodiumchloride solution), the depth of the anesthesia was assessed for each mouse by testing the tail and the hind limb pedal reflexes. Upon absence of both reflexes, mice were transcardially perfused using phosphate-buffered saline (PBS) with (for \textit{YFP in CTX} and \textit{GFAP in HC}) or without (\textit{mScarlet in PAG}) 0.4 \% heparin (Heparin-Natrium-25000, ratiopharm), and subsequently a 4 \% paraformaldehyde solution in PBS for fixation. After dissection, brains were kept in 4 \% paraformaldehyde solution in PBS for another two hours (for \textit{YFP in CTX} and \textit{GFAP in HC}) or overnight (for \textit{mScarlet in PAG}) at 4 \textcelsius.
Brains were then washed twice with PBS and stored at 4 \textcelsius\ until sectioning. For cutting, brains were embedded in 6 \% agarose in PBS and a vibratome (Leica VT1200) was used to cut 40 $\mu$m (for \textit{YFP in CTX} and \textit{GFAP in HC}) or 60 $\mu$m (for \textit{mScarlet in PAG}) coronal sections.
Immunohistochemistry was performed in 24-well plates with up to three free floating sections per well in 400 $\mu$l solution and under constant shaking. 

For \textit{YFP in CTX} and \textit{GFAP in HC}: brain sections were incubated for one hour at room temperature in 100 mM Tris-buffered glycine solution (pH 7.4). Slices were then incubated with blocking solution (10\% horse serum, 0.3 \% Triton X100, 0.1 \% Tween 20, in PBS) for one hour at room temperature. Subsequently, sections were labeled with primary antibodies at the indicated dilutions in blocking solution for 48 hours at 4~\textcelsius \:(rabbit anti-GFAP, Acris, DP014, 1:200; chicken anti-YFP, Abcam, Ab13970, 1:1000). Primary antibody solutions were washed off thrice with washing solution (0.1 \% Triton X100 and 0.1 \% Tween 20 solution in PBS) for ten minutes each. Sections were then incubated with fluorescently labeled secondary antibodies at 0.5 $\mu$g / ml in blocking solution for 1.5 hours at room temperature (goat anti-chicken Alexa-488 conjugated, Invitrogen; donkey anti-rabbit Cy3 conjugated, Jackson ImmunoResearch). Finally, sections were incubated again twice for ten minutes with washing solution and once with PBS at room temperature, prior to embedding in Aqua-Poly/Mount (Polysciences).

For \textit{mScarlet in PAG}: brain sections were incubated in blocking solution (10 \% donkey serium, 0.3 \% Triton X100, 0.1 \% Tween in 1x TBS) for two hours at room temperature. For labeling, sections were incubated for two days at 4 \textcelsius \: with rabbit anti-RFP (Biomol, 600-401-379, 1:1000) in 10 \% blocking solution in 1X TBS-T. Sections were washed thrice with washing solution for ten minutes each and then incubated with the fluorescently labeled secondary antibody at 0.5 $\mu$g / ml (donkey anti-rabbit Cy3, Jackson ImmunoResearch). Following a single wash with 1x TBS-T for 20 minutes at room temperature, sections were incubated with DAPI (Roth, 6335.1, 1:5000) in TBS-T for five minutes and eventually washed twice with 1x TBS-T. The labeled sections were embedded in embedding medium (2.4 g Mowiol, 6 g Glycerol, 6 ml ddH2O, diluted in 12 ml 0.2 M Tris at pH 8.5).

\subsubsection{Image acquisition, processing, and manual analysis}
Image acquisition was performed using a Zeiss Axio Zoom.V16 microscope, equipped with a Zeiss HXP 200C light source, an Axiocam 506 mono camera, and an APO Z 1.5x/0.37 FWD 30mm objective. Images covering 743.7 x  596.7 $\mu$m of the corresponding brain regions at a resolution of 3.7 px / $\mu$m were acquired as 8-bit images. 

To foster manual ROI annotation, these raw 8-bit images were enhanced for brightness and contrast using the automatic brightness and contrast enhancer implemented in Fiji \citep{schindelin2012fiji}. 
The corresponding image features of interest were manually annotated by Ph.D.-level neuroscientists. 
The datasets were randomly split into twelve images used for training and eight images used for evaluation.

\clearpage

\clearpage
\section{Extended performance comparison}

\subsection{Expert annotation comparison}

\begin{figure}[h!]
\includegraphics[center]{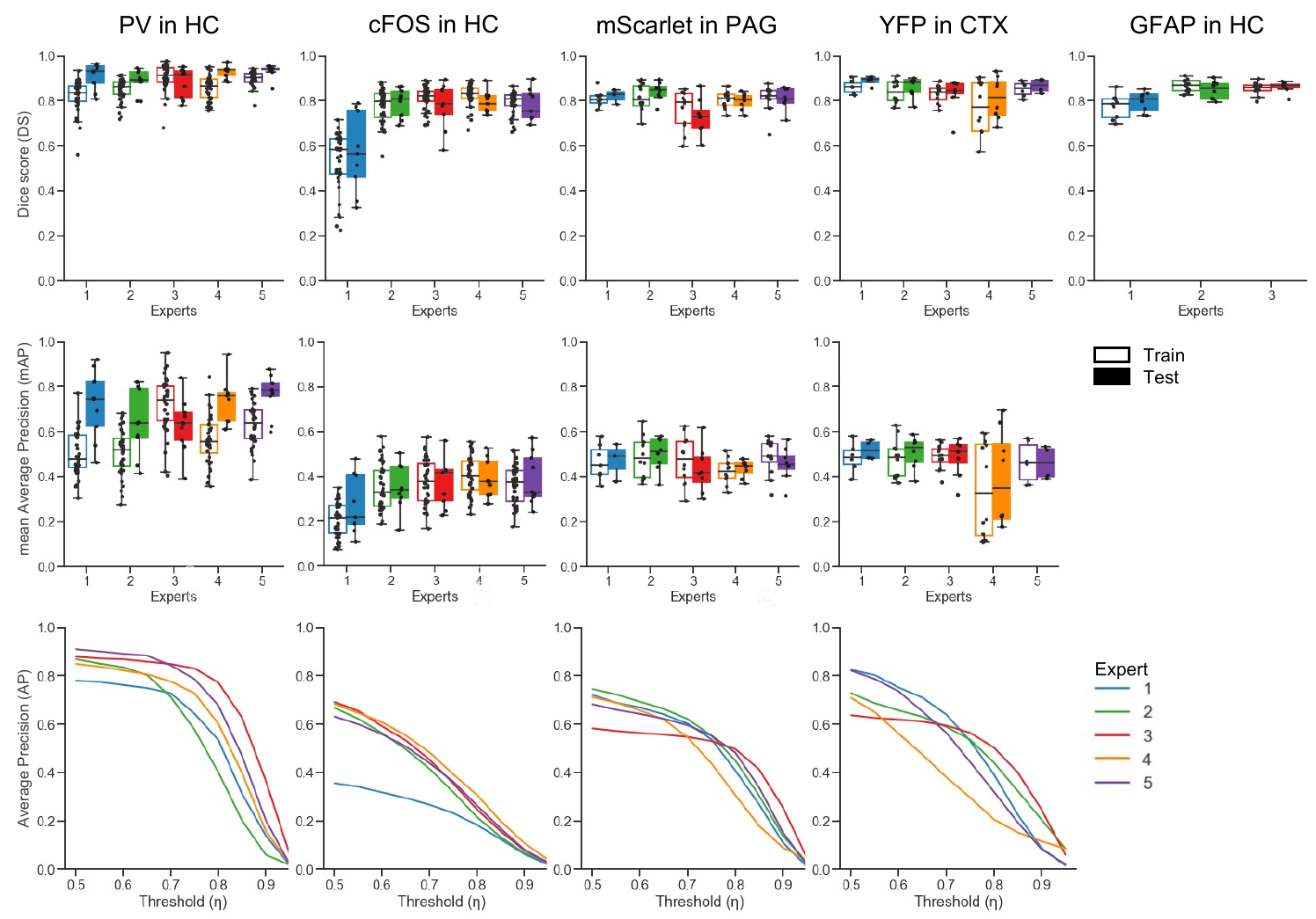}
\caption{\textbf{Expert annotation performance.} The expert annotation performance is measured by the experts' annotation similarity to the estimated ground truth (STAPLE \citep{warfield2004simultaneous}). For semantic segmentation (first row) and instance segmentation (second row) the train sets ($N=36$ images for \textit{PV in HC} and \textit{cFOS in HC}; $N=20$ for the remaining datasets) and test sets ($N=8$ images for each dataset) are depicted separately. The lines in the third row depict the mean over the Average Precision at different thresholds for all data points.
The datasets are annotated by different experts (expert IDs cannot be used as identifiers across datasets).}
\label{fig:experts}
\end{figure}

We utilize the inter-expert variation as a proxy for data ambiguity. Fig. \ref{fig:experts} depicts this variation by means of similarity scores between expert segmentations and estimated GT (derived via simultaneous truth and performance level estimation (STAPLE) \citep{warfield2004simultaneous}).
We use the dice score ($DS$, first row) for semantic segmentation and the mean Average Precision ($mAP$, second row) for instance segmentation. Both metrics exhibit a clear correlation for all experts. The performance of the experts on the train and test data is also approximately equally distributed, except for \textit{PV in HC} dataset. 
The Average Precision at different IoU-thresholds $\eta$ (third row) reveals further differences between the experts. For the \textit{mScarlet in PAG} dataset, for instance, Expert 3 has a comparatively low $AP$ at low $\eta$, but outperforms the other experts at high $\eta$. In other words, Expert 3 has a low detection performance $AP_{IoU=0.50}$ but the segmentations of the detected instances are very precise. 

\clearpage
\subsection{Method performance comparison}
To ensure that our results are robust and reproducible we repeat our experiments with different seeds. This changes the train-validation splits and weight initialization for each repetition.
In addition to the out-of-the-box \textit{cellpose*} approach \citep{stringer2021cellpose} we include fine-tuned \textit{cellpose} models and ensembles in this comparison.
The \textit{cellpose} (single) models are trained on 90/10 train-validation-splits with the default training settings from the command line interface (500 epochs, 0.2 learning rate, batch size of 8).
The \textit{cellpose} (ensemble) models are trained via five-fold cross validation and the same default training settings, resulting in an ensemble of five models similar to the \textit{deepflash2} model ensembles.
The results of three experiment repetitions for all methods are depicted in Table \ref{tab:repeat}. 

\begin{table}[htb]
	\caption{\textbf{Methods performance comparison.} Average predictive performance measured by similarity to the estimated ground truth (STAPLE \citep{warfield2004simultaneous}) on the hold-out test sets ($N=8$ images for each dataset) over three repetitions. *indicates out-of-the-box methods (not fine-tuned to the respective dataset)}
	\centering
	\scalebox{.9}{
	\begin{tabular}{llllll}
		\toprule
		   & PV in HC & cFOS in HC & mScarlet in PAG & YFP in CTX & GFAP in HC \\
		\cmidrule(r){2-6}
		Method   & \multicolumn{5}{c}{Semantic Segmentation - Mean DS (std. deviation)}            \\
		\midrule
		Otsu* & 0.101 (--) &0.033 (--) & 0.156 (--) & 0.743 (--)  & 0.600 (--)   \\
		U-Net (2019)  & 0.863 (0.010) & 0.769 (0.010) & 0.756 (0.008) & 0.850 (0.037)  & 0.762 (0.014) \\
		nnunet     & 0.891 (0.002) & 0.797 (0.000) & 0.821 (0.002) & 0.883 (0.000)  & 0.797 (0.000) \\
		deepflash2 (ours) & \textbf{0.912} (0.006) & \textbf{0.813}	(0.002) & \textbf{0.824} (0.002) & \textbf{0.886} (0.002)  & \textbf{0.809} (0.001)  \\
		\midrule
		& \multicolumn{5}{c}{Instance Segmentation - Mean mAP (std. deviation)}    \\
		\cmidrule(r){2-6}
		cellpose* & 0.541 (--) & 0.268 (--) & 0.138 (--)& 0.302 (--)& -- (--)\\
		cellpose (single) & 0.610 (0.012) & 0.329 (0.010)  & 0.415 (0.004) & 0.497 (0.014) &  -- (--) \\
		cellpose (ensemble) &0.628 (0.028) & 0.350 (0.004) & 0.431 (0.002) & 0.508 (0.010) & -- (--)   \\
		U-Net (2019)  & 0.548 (0.024) & 0.305 (0.016)  & 0.337	(0.003) & 0.452	(0.058) & -- (--)  \\
		nnunet     & 0.643	(0.004) & 0.368	(0.002) & 0.443	(0.002) & 0.528	(0.001) & -- (--) \\
		deepflash2 (ours) & \textbf{0.693} (0.007) & \textbf{0.404} (0.004) & \textbf{0.460}	(0.003) & \textbf{0.535}	(0.001) &  -- (--) \\
		\midrule
		& \multicolumn{5}{c}{Detection - $AP_{IoU=0.50}$ (std. deviation)}    \\
		\cmidrule(r){2-6}
		cellpose* & 0.701 (--) & 0.404 (--) & 0.229 (--)& 0.533 (--)& -- (--)\\
		cellpose (single) & 0.844 (0.025) & 0.662 (0.008)  & 0.666	(0.011) &  0.801 (0.027) &  -- (--) \\
		cellpose (ensemble)  & 0.851 (0.031) & 0.688 (0.015)  & 0.686 (0.009)  & \textbf{0.819} (0.018) &  -- (--) \\
		U-Net (2019)   & 0.844 (0.016) & 0.566 (0.017)   & 0.573 (0.004)  & 0.751 (0.044)  &  -- (--) \\
		nnunet & 0.825 (0.003) & 0.647 (0.011) & 0.670 (0.003)  & 0.809	(0.001) &  -- (--) \\
		deepflash2 (ours)  & \textbf{0.860}	(0.009) & \textbf{0.695} (0.003) & \textbf{0.713} (0.008) &  \textbf{0.819} (0.004) &  -- (--) \\
		\bottomrule
	\end{tabular}
	}
	\label{tab:repeat}
\end{table}

\textit{deepflash2} outperforms the other methods for both semantic and instance segmentation on all datasets. Moreover, the results show that the ensemble based methods \textit{nnunet} and \textit{deepflash2} yield very stable results (low std. deviations) across all datasets, while the U-Net of \citep{falk2019u}, based on a single model, is subject to higher performance variability. 

\begin{figure}[htb]
\includegraphics[width=0.85\linewidth, center]{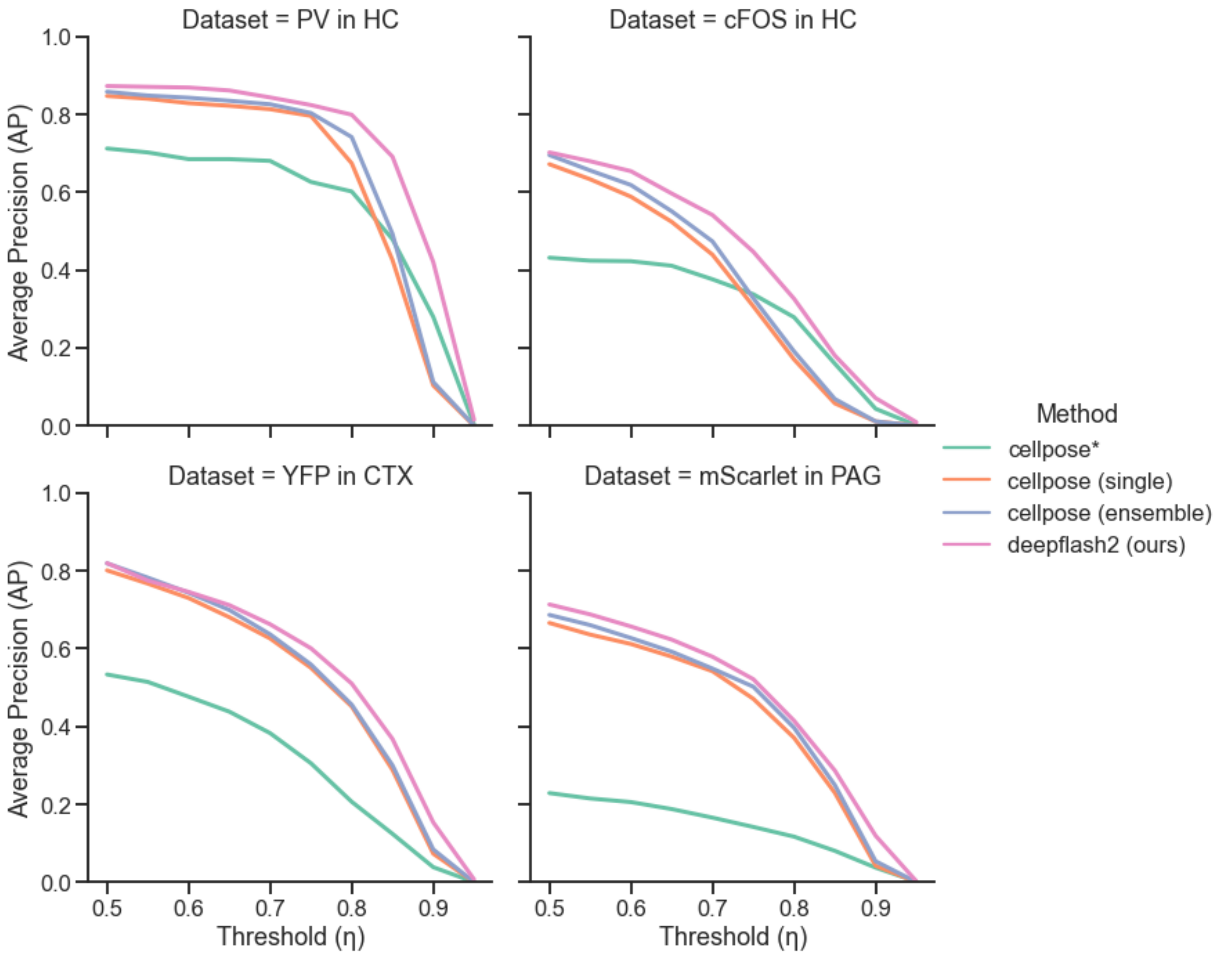}
\caption{\textbf{\textit{deepflash2} vs. \textit{cellpose}} The lines depict the instance segmentation performance using the mean of the Average Precision ($N=8$ hold-out test images for each dataset) at a certain IoU-threshold $\eta$ over 3 repetitions. The out-of-the-box model ensembles \textit{cellpose*} do not achieve competitive performance. The fine-tuned \textit{cellpose} models (single and ensemble) yield similar results to \textit{deepflash2} at low $\eta$ but constantly perform worse at higher $\eta$. 
}
\label{fig:cellpose}
\end{figure}

We also report the results of the detection task, which is commonly measured by the $AP_{IoU=0.50}$. In contrast to the $mAP$ that provides a measure for the quality of the segmentation, the $AP_{IoU=0.50}$ metric measures the ``counting'' performance of a method. That is, for instance, whether the same cell is annotated or not. The \textit{cellpose} ensemble performs on par with the \textit{deepflash2} model on the detection task on the \textit{YFP in CTX} dataset.
A more detailed analysis in Fig. \ref{fig:cellpose} reveals that the fine-tuned \textit{cellpose} models yield similar results to \textit{deepflash2} at low IoU-thresholds $\eta$ but constantly perform worse for higher thresholds. 
\clearpage

\clearpage
\section{Aleatoric and epistemic uncertainty}

\begin{figure}[!ht]
\includegraphics[center]{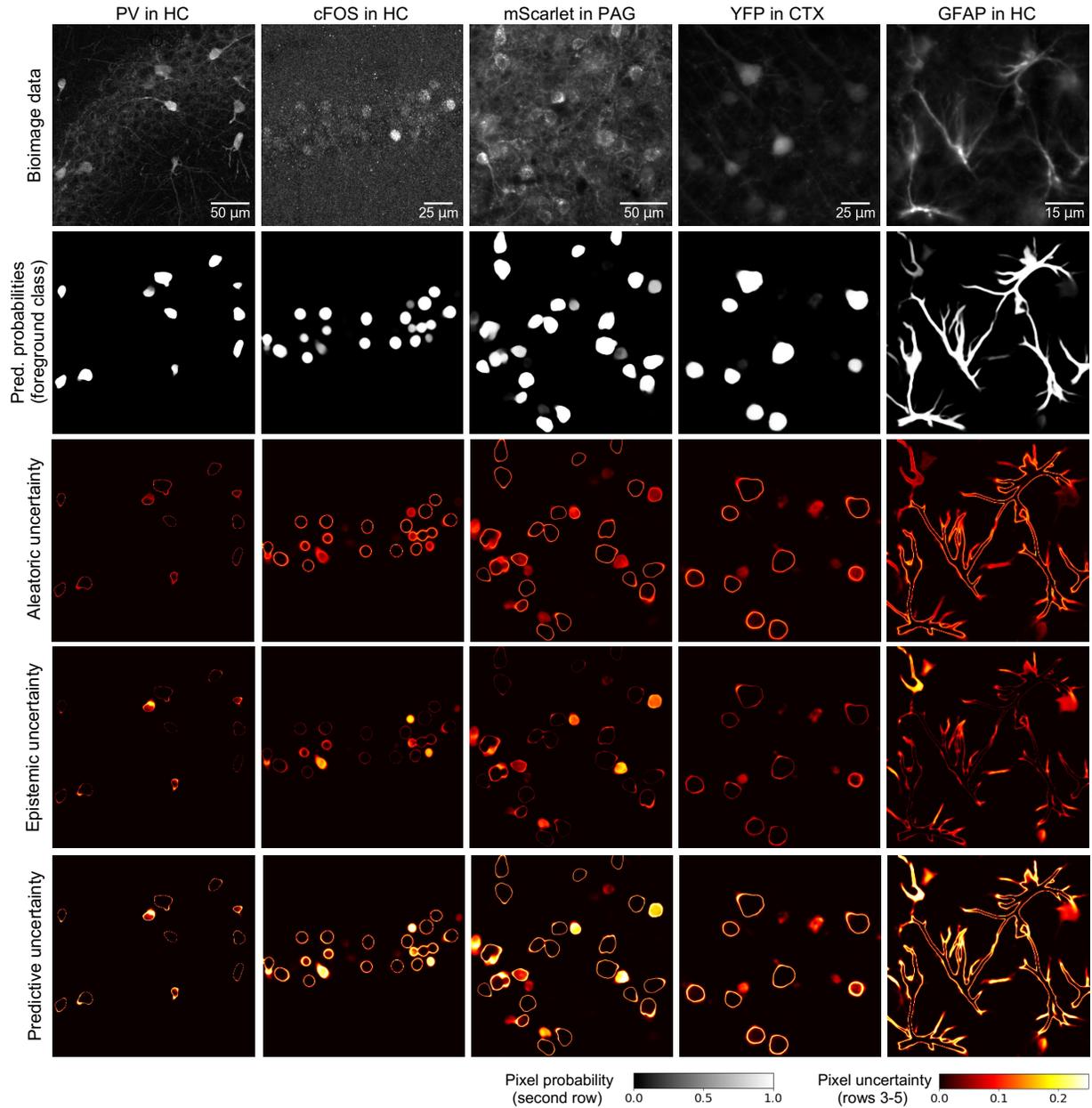}
\caption{\textbf{Visualization of uncertainty types.} Representative image sections from the test sets of five immunofluorescence imaging datasets (first row) and corresponding ensemble probability map $p(\mathbf{y}=k \mid \mathbf{X})$ for the foreground class $k$ (second row). The \textit{aleatoric} (third row) and \textit{epistemic} (fourth row) uncertainty maps are combined into the \textit{predictive} uncertainty map ${Var}_{p\left(\mathbf{y} \mid \mathbf{X},\theta \right)}$ (fifth row). \textit{Aleatoric} uncertainties foremost emerge in the border regions of the segmented cell nuclei or somata. High \textit{epistemic} uncertainties typically occur in planar areas where a clear distinction between the foreground and background classes is not feasible.
\textit{deepflash2} computes the \textit{predictive} uncertainty by default. 
The maximum pixel uncertainty is, by definition, limited to 0.25.
}
\label{fig:uncertainties}
\end{figure}

\textit{deepflash2} generates \textit{predictive} uncertainty maps that allow the quantification of data ambiguities and simplify the verification of the predictions. We show that these uncertainties can be related to the agreement of the human experts (see Fig. 2e in the main paper). The \textit{predictive} uncertainty is composed of the \textit{aleatoric} and \textit{epistemic} uncertainty (see Methods 2.2). Examples of the different uncertainty types are depicted in Fig. \ref{fig:uncertainties}. 
The \textit{aleatoric} (data or per-measurement) uncertainty can be derived from the predicted probabilities. It is low for probabilities close to zero or one, and high for probabilities around $0.5$. This results, for instance, in high \textit{aleatoric} uncertainties for the border regions of the segmented cell nuclei or somata. 
In contrast, \textit{epistemic} (model) uncertainty foremost captures the uncertainty of planar areas that are entirely ambiguous. In these areas, the models' predictions may differ considerably as a clear distinction between the foreground and background classes is not feasible. 
 %

\clearpage
\section{Extended out-of-distribution detection analysis}

\begin{figure}[h!]
\begin{center}
\includegraphics[width=.6\linewidth]{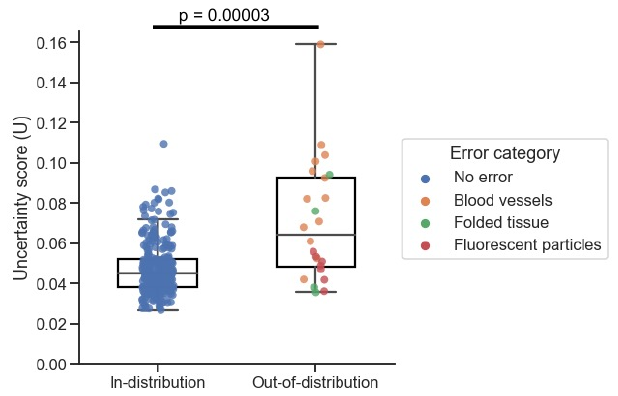}
\caption{\textbf{Uncertainty scores and out-of-distribution error categories.} Uncertainty score comparison for the \textit{cFOS in HC} out-of-distribution dataset. The dataset comprises in-distribution (No error, $N=264$) and partly out-of-distribution images (blood vessels ($N=13$), folded tissue ($N=4$), fluorescent particles ($N=7$)). 
The p-value results from a two-sided non-parametric Mann–Whitney \textit{U} test. Partly out-of-distribution images typically exhibit a higher uncertainty score.
}
\label{fig:ood}
\end{center}
\end{figure}

The quality assurance process of \textit{deepflash2} (Section 2.5 in the main paper) helps the user prioritize the review of more ambiguous images and also facilitates the detection of out-of-distribution images. Such images differ from the training data and are typically prone to erroneous predictions. In out exemplary \textit{cFOS in HC} dataset for out-of-distribution detection we differentiate between fully out-of-distribution images, e.g., images that visualize different immunofluorescent labels, and partly out-of-distribution images. The latter exhibit the same properties as the training data but also contain previously unseen structures such as blood vessels, folded tissue, or fluorescent particles (see Section \ref{sec:datasets} and Fig. \ref{fig:ood_plain}). Fig. \ref{fig:ood} shows that the uncertainty scores $U$ (Equation 5 in Section 2.2 in the main paper) of the partly out-of-distribution images are significantly higher than the uncertainty scores of the in-distribution images (No error). However, the error categories are not distributed evenly. On the one hand, blood vessels and folded tissue images cover a high and relatively wide range of uncertainty scores. On the other hand, fluorescent particle images exhibit uncertainty scores close to the median of the in-distribution images. A possible explanation is the small proportion of unseen structures (a single strongly fluorescent particle unrelated to the actual fluorescent label) in these images. Using the proposed heuristic search strategy (Fig. 2f in the main paper) such images would be detected at a later stage, however, the inclusion of such images into the bioimage analysis would possibly not impair the results.

\clearpage
\begin{sloppypar}

\bibliography{main.bib}
\end{sloppypar}